%% file: main.tex
\documentclass{article}
\usepackage{fullpage}
\usepackage{authblk}

\usepackage{soul}

\usepackage{microtype}
\usepackage{graphicx}
\usepackage{subfigure}
\usepackage{booktabs} %
\usepackage{natbib}

\usepackage[colorlinks=true,allcolors=blue]{hyperref}

\usepackage{wrapfig}
\usepackage{bbm,graphicx,bm} %
\usepackage{amsmath,amssymb,amsthm}
\usepackage{enumitem}
\usepackage{multirow}
\usepackage[all,import]{xy}
\usepackage{appendix}
\usepackage{comment}
\usepackage{color}
\usepackage{soul}
\usepackage{todonotes}
\usepackage{arydshln}
\usepackage{colortbl}
\usepackage{setspace}

\newcommand{\Cov}{\text{Cov}}

\theoremstyle{definition}
\newtheorem{definition}{Definition}

\theoremstyle{plain}
\newtheorem{proposition}{Proposition}
\newtheorem{theorem}{Theorem}

\title{Deconfounding Scores: Feature Representations for Causal Effect Estimation with Weak Overlap}
\date{}

\author[1]{Alexander D'Amour\thanks{alexdamour@google.com}}
\author[2]{Alexander Franks\thanks{afranks@pstat.ucsb.edu}}

\affil[1]{Google Research, Cambridge, MA, USA}
\affil[2]{Department of Statistics and Applied Probability,
\authorcr University of California, Santa Barbara, Santa Barbara, CA, USA}

\begin{document}
\maketitle

\begin{abstract}
A key condition for obtaining reliable estimates of the causal effect of a treatment is overlap (a.k.a. positivity): the distributions of the features used to perform causal adjustment cannot be too different in the treated and control groups. In cases where overlap is poor, causal effect estimators can become brittle, especially when they incorporate weighting. To address this problem, a number of proposals (including confounder selection or dimension reduction methods) incorporate feature representations to induce better overlap between the treated and control groups. A key concern in these proposals is that the representation may introduce confounding bias into the effect estimator. In this paper, we introduce deconfounding scores, which are feature representations that induce better overlap without biasing the target of estimation. We show that deconfounding scores satisfy a zero-covariance condition that is identifiable in observed data. As a proof of concept, we characterize a family of deconfounding scores in a simplified setting with Gaussian covariates, and show that in some simple simulations, these scores can be used to construct estimators with good finite-sample properties. In particular, we show that this technique could be an attractive alternative to standard regularizations that are often applied to IPW and balancing weights.
\end{abstract}

\onehalfspacing
\section{Introduction}

Estimating the causal effect of a treatment on an outcome from observational data is one of the central tasks in causal inference.
When treatments are not randomly assigned, observed correlations need to be adjusted to recover the causal relationship between treatment and outcome that we would observe if we were able to intervene on the population and assign treatments directly. 
A standard approach is to adjust by conditioning confounders, or pre-treatment variables that satisfy conditional ignorability or the backdoor criterion \citep{imbens2015causal,hernan2020causal,pearl2009causality}.
Conditioning on these features explains away non-causal covariation between the treatment and the outcome that is driven by third variables that affect them both. 

Importantly, many estimators that incorporate confounder adjustment only yield reliable estimators if there is substantial overlap in the distributions of these features in the treated and control groups.
In particular, if overlap in confounders is poor, these estimators can exhibit large variance and substantial finite-sample departures from their asymptotic behavior.
This is particularly true of so-called doubly-robust estimators that incorporate an element of importance weighting \citep{kang2007demystifying}, and which serve as an important interface for applying machine learning models to causal effect estimation problems \citep[see, e.g.,][]{athey2019mlreview}. 
Unfortunately, poor overlap is likely to occur in many moderns settings where it is necessary to adjust for a large number of features to plausibly address confounding \cite{d2017overlap}.

In this work, we present some preliminary results on a feature representation approach to improve the finite-sample behavior of causal effect estimators under poor overlap.
The key idea is that, given a set of features $X$ that are sufficient to address confounding, there exist many non-invertible feature representations $d(X)$ that also yield unbiased treatment effect estimates.
We call these reductions \emph{deconfounding scores}.
Deconfounding scores have appeared before in the causal inference literature, including the propensity score \citep{rosenbaum1983central} and the prognostic score \citep{hansen2008prognostic}, but here we show that a more general family of such scores exists.
Importantly, any deconfounding score is sufficient to adjust for confounding in estimating a causal effect, and it turns out that most deconfounding scores exhibit better overlap than raw features, and can thus be used to construct more stable estimators.

As a proof of concept, we explore how deconfounding scores can be used to construct more stable weighting-based estimators of causal effects, and show some promising results in a simplified setting.
Specifically, we analytically characterize a family of deconfounding scores in a setting where features are Gaussian, and the outcome and treatment assignment follow generalized linear models.
In this case, the family can be elegantly characterized in terms of a hyperbolic constraint, where the propensity score and prognostic score constitute extreme values of the set.
Based on this insight, we propose parameterizing deconfounding scores in this setting in terms of their similarity to the propensity score, and develop a corresponding set of weighting estimators that we test in simulation.
We find that using deconfounding scores to define alternative weights in inverse probability weighting (IPW) and doubly-robust estimators can yield substantial improvements in mean squared error, and that this approach is a desirable alternative to applying strong regularization or clipping to weights obtained from raw features. 

We make the following contributions:
\begin{itemize}
\item We define general deconfounding scores, and specify an identifiable condition for computing them. 
\item We analytically characterize a family of deconfounding scores $d(X)$ in a simple setting, and define a corresponding family of estimators.
\item In simulations, we show that replacing propensity score weighting with weighting derived from the deconfounding score can result in substantial mean squared error gains, in contrast to applying strong regularization or clipping to propensity scores.
\end{itemize}

\section{Preliminaries}
\subsection{Problem Setup and Notation}
We consider an observational study where we wish to learn the population average causal effect of some binary treatment $T$ on some observed outcome $Y$.
We model our observations as independent and identically distributed (i.i.d.) draws from a population, where each unit has potential outcomes $(Y(0), Y(1))$, and a set of features $X$ that may influence treatment and/or potential outcomes. 
Our effect of interest, the Average Treatment Effect (ATE), can be written as:
$$
\tau^{ATE} := E[Y(1) - Y(0)].
$$

For each unit, the outcome $Y$ that we observe is determined by the treatment assignment $T$:
$$
Y = T \cdot Y(1) + (1 - T) \cdot Y(0).
$$

In this setting, the following \emph{strong ignorability} conditions are sufficient to identify the ATE \cite{imbens2015causal}:
\begin{align}
\label{eq:strong ignorability}
(Y(0), Y(1)) \perp T \mid X \quad \text{and} \quad 0 < P(T = 1 \mid X) < 1 \textrm{  a.s.}.
\end{align}
These are the unconfoundedness and overlap conditions, respectively.
Unconfoundedness stipulates that, conditional on $X$, treatment assignment is as-good-as-random.
The overlap condition stipulates that units with each value of the covariates are at risk of receiving both values of the treatment.
Under these assumptions, $\tau^{ATE}$ is non-parametrically identified by the following functional, which we call the \emph{statistical estimand}: 
\begin{equation}
\label{eq:statistical estimand}
\tau^{ATE} = E[E[Y \mid X, T = 1] - E[Y | X, T = 0]].
\end{equation}
Specifically, unconfoundedness implies that this equality is valid, while the overlap condition ensures that the both conditional expectations can be estimated without parametric assumptions.

For our discussion, two conditional expectation functions play an important role.
We call the conditional expectation of treatment assignment the \emph{propensity score} and denote it $e(X) := E[T \mid X]$.
We call the the conditional expectation of the outcome under each treatment assignment \emph{prognostic scores}
\footnote{Here, we slightly modify the terminology of \cite{hansen2008prognostic}, who only referred to $m_0(X)$ as the prognostic score.}
and denote them by $m_t(X) := E[Y \mid X, T = t]$ for $t = 0, 1$.

Finally, for the remainder of the paper, we will refer to ``overlap'' more generally than to refer to the overlap condition in \eqref{eq:strong ignorability}.
In particular, we will use ``overlap'' to mean the extent to which extreme propensity scores close to 0 or 1 occur with low probability.
We will say overlap is stronger when this probability is low, and overlap is poor when this probability is high.

\subsection{Weighting Estimators and the Importance of Overlap}

In this paper, we focus on how poor overlap affects the reliability of causal effect estimators and how to mitigate these effects.
In particular, we will focus on estimators that incorporate importance weights, because for these estimators, the importance of overlap is directly apparent.
Two such estimators are vanilla inverse probability weighting (IPW), which reweights observed outcomes by the estimated probability of the treatment received:
\begin{align}
\label{eq:IPW}
\hat \tau_{IPW}^{ATE} := \hat E\left[\left(\frac{T}{\hat e(X)} - \frac{(1-T)}{(1-\hat e(X))}\right)Y\right],
\end{align}
and doubly robust methods that combine the propensity score $e(X)$ and the prognostic scores $m_t(X)$. Although there are a number of different methods of combination, we focus on the augmented IPW (AIPW) estimator with the following form:
\begin{align}
\label{eq:AIPW}
&\hat \tau_{AIPW}^{ATE} :=  \hat E[\hat m_1(X)] - \hat E[\hat m_0(X)] + \hat E\left[\left(\frac{T}{\hat e(X)} - \frac{(1-T)}{(1-\hat e(X))}\right)(Y - \hat m_T(X))\right].
\end{align}
Intuitively, the approach uses inverse-propensity weighting to reweight the residuals from a plug-in estimate of \eqref{eq:statistical estimand}.

Doubly robust methods are particularly important for applications of machine learning to causal effect estimation because they are able to reduce some of the bias induced by regularization \cite{van2011targeted,10.1111/ectj.12097}. 
For example, in \eqref{eq:AIPW}, when the estimate of the prognostic models $\hat m_t(X)$ is biased (e.g., due to regularization), the weighted residual term can mitigate this bias, even when the propensity score is itself estimated with regularization.

The form of these estimators immediately reveals why they are sensitive to poor overlap.
In cases where the features $X$ are highly predictive of treatment assignment, very small values can occur in denominators and dominate the estimate.

\section{Related Work}
A number of approaches attempt to improve the reliability of causal effect estimators by modifying the features $X$.
A key insight, which has driven a number of these approaches, is that, for a set of features $X$ that satisfy unconfoundedness, only those portions of $X$ that are predictive of \emph{both} treatment assignment and outcome are relevant for confounding.
Furthermore, while portions of $X$ that predict the outcome $Y$ can be helpful for improving precision, portions of $X$ that predict treatment assignment alone only reduce overlap.
Thus, these approaches attempt to remove information in $X$ that is predictive of the treatment assignment $T$.

In the statistics and epidemiology literatures, it has long been considered a best practice to remove instruments, or features that are solely predictive of treatment assignment \citep[see, e.g.,][]{vanderweele2019principles,rubin1997estimating}.
In some cases this practice has been formalized in feature selection algorithms that select features based on their observed relationship to outcomes \citep{van2010collaborative,shortreed2017outcome} or based on graphical criteria \citep{rotnitzky2019efficient}.
Beyond feature selection, more general representation learning approaches that obtain functional reductions of $X$ have been proposed in the statistics \citep{Luo2017} and machine learning literatures \citep{johansson2016learning,shalit2017estimating}.
These methods attempt to obtain a function of $X$ that is sufficient only for predicting outcomes, while shedding all other information about treatment assignment.
Here, our representation learning approach is distinct: instead of focusing on sufficiency for outcome prediction, we focus on minimizing an estimate of bias, which yields a broader family of valid representations.

Constructing more reliable weighted causal effect estimators using so-called balancing weights is also an active area of research.
In particular, a number of recently-proposed approaches consider replacing the inverse propensity weights in \eqref{eq:IPW} and \eqref{eq:AIPW} with modified weights that are constrained to balance a class of potentially outcome-relevant functions or sufficient statistics.
Some of these approaches include entropy balancing \cite{hainmueller2012entropy,zhao2017entropy}, approximate balancing weights \cite{wang2017minimal}, and approximate residual balancing \cite{athey2018approximate}, along with analogous approaches in the reinforcement leraning off-policy evaluation literature \citep[e.g., ][]{kallus2018balanced}.
When overlap is poor, these estimators achieve variance reductions relative to standard IPW methods in two ways: first, by restricting the set of functions to be balanced to a class that potentially has better overlap, and second, by applying regularization directly to the weights rather than to the probability of treatment. 

The specific weighting estimators that we derive using deconfounding scores have a similar flavor, but we specify the balancing conditions in a different way that is especially amenable to the poor overlap setting. 
In particular, our estimators are specified to balance a single function $d(X)$, which gives more fine-grained control over the family of functions that the resulting weights will balance.
We discuss this distinction in more detail in Section~\ref{sec:results}.
We note, however, that these approaches are not mutually exclusive; deconfounding scores and balancing weights could be combined in a number of ways.

\section{Deconfounding Scores}
In this section, we describe deconfouding scores, a tool that we use to develop treatment effect estimators for studies with poor overlap.
Although we believe deconfounding scores have a number of use cases, we will focus on using them to develop weights that can be used in IPW and AIPW estimators.
Our end goal will be to construct a set of reduced propensity scores $e_d(X)$ that can be substituted into the expressions \eqref{eq:IPW} and \eqref{eq:AIPW}.

\subsection{Beyond Propensity and Prognostic Scores}

At a high level, deconfounding scores $d(X)$ are functional reductions of the covariates that discard some information, but can still be used to obtain unbiased ATE estimates.
\begin{definition}
\label{def:deconfounding}
A function of covariates $d(X)$ is a deconfounding score $X$ if and only if it satisfies
\begin{align}
\tau^{ATE} = \tau_d^{ATE} := E\{E[Y \mid d(X), T = 1] - E[Y \mid d(X), T = 0]\}.
\end{align}
\end{definition}
Deconfounding scores are not a new idea: both the propensity score $e(X)$ and the prognostic scores $m_t(X)$ are examples of deconfounding scores.
However, the class of deconfounding scores is more general.
In our case, we require a deconfounding score that is neither a propensity score nor a prognostic score: we want to shed some of the information about treatment assignment contained in $e(X)$, but we want to incorporate information beyond what is contained in $m_t(X)$ to retain the double robustness property of the AIPW estimator.

To construct a weighting estimator from a deconfounding score, we can compute the reduced propensity score $e_d(X) := P(T = 1 \mid d(X))$ and weight by that quantity.
Importantly, for any deconfounding score that is not a balancing score, $d(X)$ will be insufficient for $T$, and thus will induce less extreme propensity scores $e_d(X)$ on average compared to the full propensity score $e(X)$.

\subsection{Computing Deconfounding Scores}

From the expression in Definition~\ref{def:deconfounding}, it is not obvious how one would compute a deconfounding score $d(X)$.
By contrast, propensity scores $e(X)$ and prognostic scores $m_t(X)$ are defined by sufficiency relations for $T$ and $Y$: conditional on the propensity or prognostic score, $T$ or $Y(t)$ is independent of $X$, respectively.
Thus, $e(X)$ and $m_t(X)$ can be computed by solving standard regression problems.
Computing general deconfouding scores requires a different approach.

Here, we develop one approach by analyzing the bias that is incurred conditioning on a reduction $d(X)$ instead of the full set of covariates $X$ (for which unconfoundedness is assumed to hold). 
The bias can be expressed as follows.

\begin{proposition}[Reduction Bias Expression]
\label{prop:bias formula}
For any function of covariates $d(X)$, the bias of the statistical estimand $\tau^{ATE}_d$ is 
\begin{align}
\label{eq:theory equation}
\tau^{ATE} - \tau_d^{ATE} = E\left[\frac{\Cov(Y(1), T \mid d(X))}{e_d(X)} +\frac{\Cov(Y(0), T \mid d(X))}{1-e_d(X)} \right],
\end{align}
where $e_d(X) := P(T = 1 \mid d(X))$
\end{proposition}

The proof is included in the appendix.
This expression is intuitive and shows precisely how unexplained correlation between the treatment assignment $T$ and potential outcomes contribute to bias in effect estimation.
We note that \eqref{eq:theory equation} is a fully general expression for the bias due to unobserved confounding when conditioning on $d(X)$, and without assumptions on $X$, this quantity is generally unidentified.

Importantly, when $X$ satisfies unconfoundedness and the outcome and propensity models are estimable, this bias is identifiable.
Thus, it can be translated into a concrete constraint derived from observables.
Conveniently, this constraint can be written in terms of the propensity and prognostic scores.

\begin{proposition}[Reduction Bias Identification]
When $X$ satisfies unconfoundedness, the bias of the statistical estimand $\tau^{ATE}_d$ is identified by
\begin{align}
\label{eq:empirical bias}
\tau^{ATE} - &\tau_d^{ATE} = E\left[\frac{\Cov(m_1(X), e(X) \mid d(X))}{e_d(X)} + \frac{\Cov(m_0(X), e(X) \mid d(X))}{1-e_d(X)} \right]
\end{align}
\begin{proof}
Note that, under unconfoundedness, $\Cov(Y(t), T \mid X) = 0$ for all $X$.
Letting $m_t(X) := E[Y(t) \mid X]$,
\begin{align*}
\Cov(Y(t), T \mid d(X)) &=E[ \Cov(Y(t), T \mid X) \mid d(X) ] +
\Cov( E[Y(t) \mid X], E[T \mid X] \mid d(X) )\\
&= \Cov( m_t(X), e(X) \mid d(X)). \qquad \qedhere
\end{align*}

\end{proof}
\end{proposition}
This suggests that one approach to deriving deconfounding scores is to estimate outcome and propensity score models, and then set the empirical version of \eqref{eq:empirical bias} to zero.
We consider a case where this can be done analytically in the next section.

\section{Deconfounding Scores with Gaussian Covariates and Linear Sufficient Statistics}

In this section, we analytically explore the properties of deconfounding scores in a simplified setting. 
The setting here is meant to serve as a proof of concept, and where relevant, we highlight where results would change as we relax these strong assumptions.

To simplify our estimation task further, we focus on estimating the average treatment effect on the treated, or ATT, defined as
$$
\tau^{ATT} := E[Y(1) - Y(0) \mid T = 1]
$$
Note that the average treatment effect $\tau^{ATE}$ can be written as a weighted sum of $\tau^{ATT}$ and the analogously defined average treatment effect on the control $\tau^{ATC}$.
Under unconfoundedness, $\tau^{ATT}$ is identified by
$$
\tau^{ATT} = E[Y^{obs} \mid T = 1] - E[m_0(X) \mid T = 1].
$$
Importantly, this is only a function of the control prognostic score $m_0(X)$.
Thus, to obtain deconfounding scores for the ATT, we only need to set the second term of \eqref{eq:empirical bias} to be equal to zero.
The exercise in this section could be repeated for $m_1(X)$ to obtain estimators for $\tau^{ATC}$, and thus $\tau^{ATE}$.

We assume that the we have $p$ normally distributed observed confounders, $X \sim N(0, \Sigma)$, a binary treatment $T \in {0, 1}$, and continuous potential outcomes.  Further we assume that the expectation of the control outcome and the propensity score are
\begin{equation}
\label{eqn:model_def}
E[Y(0) \mid X] = m_0(X) = f_Y(\mathbf{\alpha}'X);\qquad
E[T \mid X] = e(X) = f_T(\mathbf{\beta}'X)
\end{equation}
respectively.  We consider deconfounding scores of the form
\begin{align}
d_\gamma(X) &= \gamma'X.
\end{align}
Without loss of generality we assume $\gamma$, $\alpha$ and $\beta$ are all unit vectors in $\mathbb R^p$, $\alpha'\beta >= 0$ and $f_Y: \mathbb R \rightarrow \mathbb R$ and $f_T: \mathbb R \rightarrow [0, 1]$. Our goal is to characterize the set of vectors $\gamma$ for which Equation \ref{eq:empirical bias} is equal to zero.  Importantly, because $X$ are Gaussian,
\begin{enumerate}
\item For all $X$, 
\begin{align*}
\Cov(\alpha'X, \beta'X \mid \gamma'X) = \alpha'\Sigma\beta - (\alpha' \Sigma \gamma) (\gamma' \Sigma \gamma)^{-1} (\gamma' \Sigma \beta).
\end{align*}
\item For any function $f_Y$ and $f_T$ (even non-linear), we have
\begin{align*}
\Cov(\alpha' X, \beta'X | \gamma'X) = 0 \implies \Cov(m_0(X), e(X) | d(X)) = 0.
\end{align*}
\end{enumerate}
Hence, we can characterize a family of deconfounding scores by finding the set of vectors $\gamma$ for which $\text{Cov}(\alpha' X, \beta'X | \gamma'X)=0$.  For simplicity, and without loss of generality, we assume that $\Sigma=I$, so that the set of deconfounding scores can be expressed as
\begin{equation}
\label{eq:gaussian constraint}
\{\gamma'X : \alpha'\gamma\gamma'\beta = \alpha'\beta\}
\end{equation}

This constraint leads to the following theorem:
\begin{theorem}
\label{thm:hyperbola}
All unit-length vectors $\gamma$ that satisfy \eqref{eq:gaussian constraint} can be expressed as
\begin{align*}
\gamma = w_1\left( \frac{(\alpha + \beta)}{\sqrt{2 + 2\alpha'\beta}}\right) + w_2\left(\frac{(\alpha - \beta)}{\sqrt{2 - 2\alpha'\beta}}\right) +
\sqrt{1-w_1^2 - w_2^2}\cdot n
\end{align*}

where $n \in \text{Null}(Span[\alpha, \beta])$ and $w_1$ and $w_2$ satisfy the equation $\left(\frac{\alpha'\beta + 1}{2\alpha'\beta}\right)w_1^2 + \left(\frac{\alpha'\beta - 1}{2\alpha'\beta}\right)w_2^2 = 1$.
\end{theorem}

\begin{figure*}[t!]
\includegraphics[width=\textwidth]{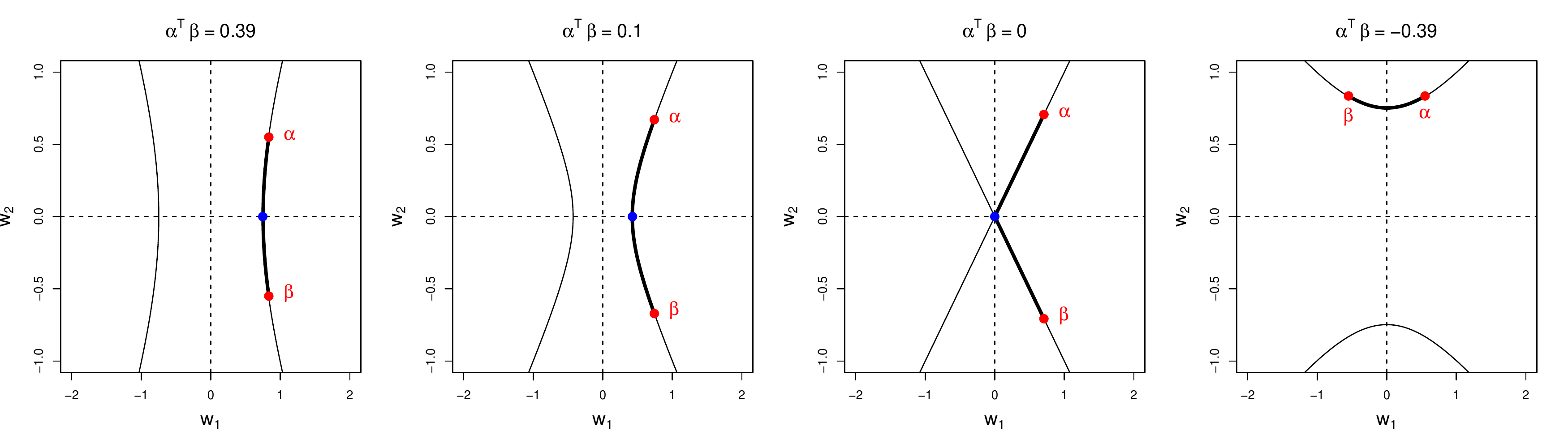}
\caption{The projection of $\gamma$ onto the space spanned by $\alpha$ and $\beta$ lies on a segment of a hyperbpola (bold black line) whose endpoints correspond to $\gamma=\alpha$ and $\gamma=\beta$.  The shape of the hyperbola depends on the inner product $\alpha'\beta$.  When $\alpha'\beta$ changes sign, the orientation of the hyperbola changes.  \label{fig:hyperbola}}
\end{figure*}

To interpret this result, we note that the space of coordinates $(w_1, w_2)$ that yield valid unit-length values of $\gamma$ correspond to a segment of a hyperbola that lies on the subspace spanned by the prognostic score and propensity score coefficient vectors $\alpha$ and $\beta$ (see Figure \ref{fig:hyperbola}). Here, $w_2$ controls the position of the projection of $\gamma$ on the continuous sheet of the hyperbola. When we move $w_2$ along its valid range $\left[-\sqrt{\frac{1-\alpha'\beta}{2}}, \sqrt{\frac{1-\alpha'\beta}{2}}\right]$, setting $w_2$ to its minimum value implies that $\gamma = \alpha$ so that $d(X) = m_0(X)$, whereas setting $w_2$ to its maximum value implies that $\gamma = \beta$ so that $d(X) = e(X)$.  $w_2 = 0$ implies $\gamma$ is equiangular to $\alpha$ and $\beta$, that is $\alpha'\gamma = \beta'\gamma$.
For points on the interior of $w_2$'s range, there is an equivalence class of $\gamma$'s: the projection of $\gamma$ onto $\mathrm{span}\{\alpha, \beta\}$ is constrained, but orthogonal component of $\gamma$ is unconstrained.
We interpret $w_2$ as a scalar parameter that controls the similarity of the deconfounding score to the propensity and prognostic scores.  
Since setting $w_2$ to its minimum value recovers the prognostic score, $\beta'X$, we conjecture that the prognostic score is the overlap maximizing deconfounding score.

\section{Experiments}
\label{sec:results}
To examine how our analytical results from the last section translate to estimation, we run several simulations where we examine the performance of the ATT analogues of the IPW \eqref{eq:IPW} and AIPW \eqref{eq:AIPW} estimators with $e(X)$ replaced with $e_d(X) = P(T = 1 \mid d(X))$ (we include full specifications of these ATT estimators in the appendix).

We state some of our findings here with the caveat that this setting is highly simplified.
First, replacing the vanilla propensity score weights with reduced propensity score weights is highly effective at lowering MSE.
Second, as a variance reduction technique, our approach dominates clipping and over-regularizing the propensity scores, which are two common techniques for mitigating extreme inverse propensity weights.
This is a direct result of the fact that our technique is designed to reduce variance without incurring bias.
Finally, we see especially strong performance for propensity weights computed with respect to the prognostic score $e_m(X) = P(T = 1 \mid m_0(X))$.

\subsection{Simulation Setup and Inference}
Using the model given in equation \ref{eqn:model_def}, we set $f_Y$ to the identity and $f_T$ to the inverse logit function.
We generate $n$ iid triples $([Y(0), Y(1)], T, X)$ according to
\begin{align}
X &\sim N_p(0, I)\\
T &\sim \text{Bern(e(X))}; \quad \label{eqn:prop_model} e(X) = \text{logit}^{-1}(\beta_0 + s_TX'\beta)\\
Y(t)  &\sim N\left(\alpha_0 + s_YX'\alpha + \tau t, \sigma^2\right)
\end{align}

Here, $s_Y$ and $s_T$ are the scale in the outcome model and propensity score model respectively. $s_Y$ corresponds to the signal-to-noise ratio for the outcome model, while $s_T$ controls overlap; higher values of $s_T$ correspond to more extreme propensity scores, and poorer overlap.
We consider high-dimensional settings where the aspect ratio $p/n \in [0.2, 2]$, so $\alpha$ and $\beta$ need to be estimated with strong regularization.
Thus, this is a setting where AIPW estimators are meant to shine.

\begin{figure*}[t!]
\centering

    \centering 
        \includegraphics[width=\textwidth]{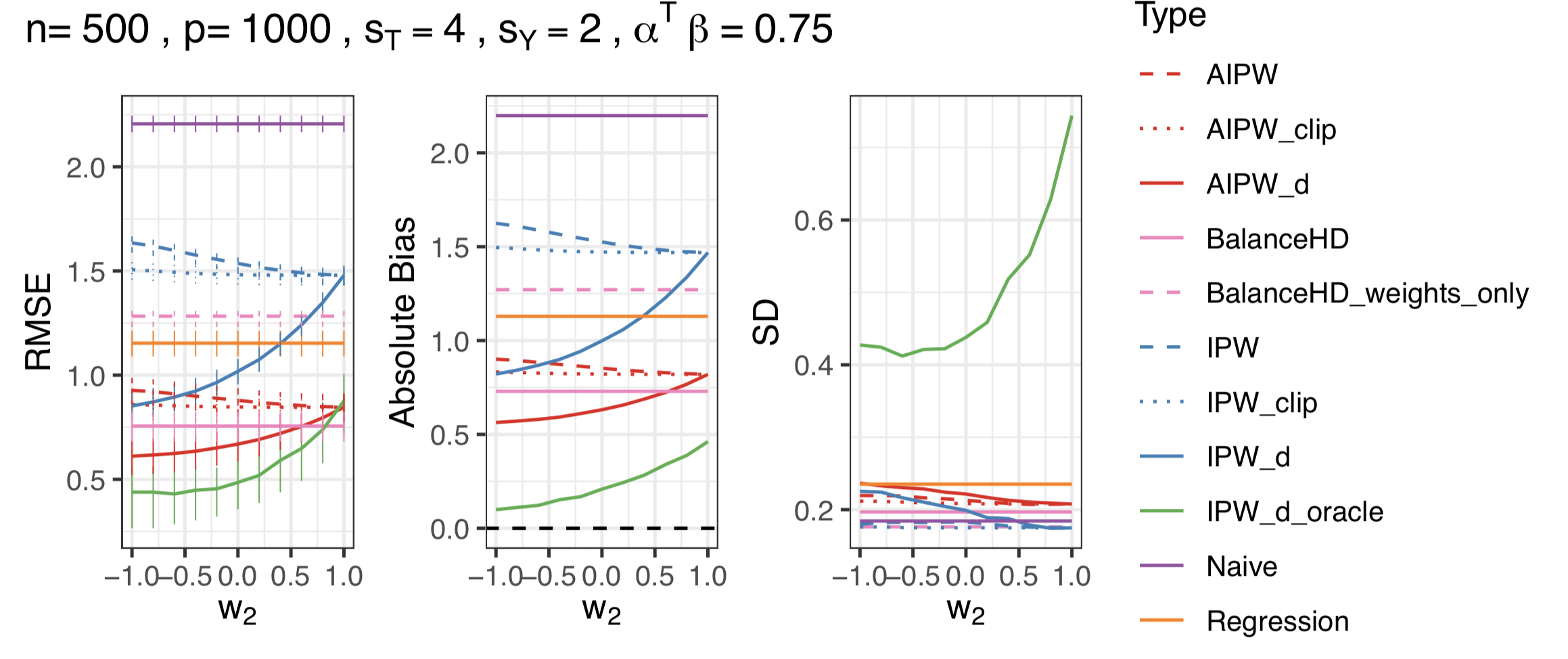}
        \label{fig:bv_est}
\caption{\textbf{RMSE, bias and standard deviation for baselines and estimators based on deconfounding scores.}  Lasso regression is used for both the outcome and propensity score models in the figures above.  IPW-d and AIPW-d utilize propensity scores estimated using our novel deconfounder score, whereas IPW and AIPW use standard regression estimates of the propensity score, with regularization chosen to match the propensity-score variance of $e_d$ for each value of $w_2$. For $w_2=1$, the propensity score regularization is determined by cross-validation. We also include clipping estimators which clip propensity scores to be in the interval [0.1, 0.9].  \label{fig:bias_var}}
\end{figure*}

We compare a variety of estimators for the ATT, which utilize estimates of $\hat m_0(X)$ and $\hat e(X)$ in different ways.
As baselines, we report the root mean squared errors for the na\"ive estimator (difference in observed means), a regression estimator that estimates \eqref{eq:statistical estimand} by plug-in, the IPW estimator \eqref{eq:IPW} and the AIPW estimator \eqref{eq:AIPW}. We compare these to IPW and AIPW estimators where the propensity score weights are replaced with reduced propensity scores $e_d(X)$, for various settings of $d(X)$.  

As a strong baseline, we also compare to the approximate residual balancing technique of \citet{athey2018approximate}, which we denote by its R package name balanceHD.
\footnote{https://github.com/swager/balanceHD}
balanceHD resembles the AIPW estimator, but replaces inverse probability weights with weights balance the means of the features in the treated and control groups; this has the effect of balancing the class of all linear functions of $X$.
This method was specifically designed to operate in this high-dimensional linear setting, but has been observed to have weaker performance in low overlap settings.

We denote our deconfounding score estimators as IPW-d(w) and AIPW-d(w) respectively, where $w$ is a normalized version of the parameter $w_2$ from Theorem~\ref{thm:hyperbola}.
$w$ quantifies the similarity of the deconfounding score to both the propensity score and the prognostic score on a scale from $-1$ to $1$.
Here, $w=1$ indicates $d(X) = e(X)$, and $w=-1$ indicates $d(X) = m_0(X)$.  $w=0$ corresponds to a deconfounding score that is equally correlated with the prognostic and propensity scores.  Note that by definition AIPW-d(1) and IPW-d(1) are equivalent to vanilla $\text{AIPW}$ and $\text{IPW}$ respectively.
We discuss how we compute $e_d(X)$ given a prognostic score $d(X)$ in the appendix.
In all cases, we estimate the prognostic score, $\hat m_0(X)$, using regularized regression (ridge or lasso). 

For all simulations we take $n=500$ and consider lower dimensional ($p=100$) and higher dimensional ($p=1000$) feature sets.  We also consider high overlap setting, $S_T = 1$ and low overlap settings $S_T = 4$.  
When $S_T = 1$  we have strong overlap, with e(X) generally between $0.2$ and $0.8$, whereas for $S_T=4$ we have weak overlap (the majority of observations $e(X)$ are outside of $[0.1, 0.9]$).
We also consider two values of $S_Y$, which controls variance explained by covariates, with larger values impliying a higher singal-to-noise regime.
We consider estimating the regression models using either ridge regression (``R") or lasso regression (``L"), with regularization parameters that are estimated via cross validation implemented in the \texttt{glmnet} R package \citep{friedman2010regularization,r}.
\footnote{
Following balanceHD, when the models are used for directly predicting outcomes, we use the \texttt{1se} rule implemented in \texttt{glmnet}. When model coefficients are used for computing deconfounding scores, we use the \texttt{lambda.min} rule instead.
}
In our simulations, the true outcome models are sparse, so the lasso regression is expected to provide a better outcome model fit than ridge regression.
We report averages across 100 runs of each setting.

\begin{table*}[t]
\centering
\tiny
\begin{tabular}{l|cccc|cccc|cccc|cccc|}
\hline
\multicolumn{1}{c|}{ } & \multicolumn{8}{c|}{Lower dimension} & \multicolumn{8}{c|}{Higher Dimension} \\
\cline{2-9} \cline{10-17}
\multicolumn{1}{c|}{ } & \multicolumn{4}{c|}{High Overlap} & \multicolumn{4}{c|}{Low Overlap} & \multicolumn{4}{c|}{High Overlap} & \multicolumn{4}{c|}{Low Overlap} \\
\cline{2-5} \cline{6-9} \cline{10-13} \cline{14-17}
\multicolumn{1}{c|}{ } & \multicolumn{2}{c|}{High SNR} & \multicolumn{2}{c|}{Low SNR} & \multicolumn{2}{c|}{High SNR} & \multicolumn{2}{c|}{Low SNR} & \multicolumn{2}{c|}{High SNR} & \multicolumn{2}{c|}{Low SNR} & \multicolumn{2}{c|}{High SNR} & \multicolumn{2}{c|}{Low SNR} \\
\cline{2-3} \cline{4-5} \cline{6-7} \cline{8-9} \cline{10-11} \cline{12-13} \cline{14-15} \cline{16-17}
\multicolumn{1}{c|}{ } & \multicolumn{1}{c|}{R} & \multicolumn{1}{c|}{L} & \multicolumn{1}{c|}{R} & \multicolumn{1}{c|}{L} & \multicolumn{1}{c|}{R} & \multicolumn{1}{c|}{L} & \multicolumn{1}{c|}{R} & \multicolumn{1}{c|}{L} & \multicolumn{1}{c|}{R} & \multicolumn{1}{c|}{L} & \multicolumn{1}{c|}{R} & \multicolumn{1}{c|}{L} & \multicolumn{1}{c|}{R} & \multicolumn{1}{c|}{L} & \multicolumn{1}{c|}{R} & \multicolumn{1}{c|}{L} \\
\cline{2-2} \cline{3-3} \cline{4-4} \cline{5-5} \cline{6-6} \cline{7-7} \cline{8-8} \cline{9-9} \cline{10-10} \cline{11-11} \cline{12-12} \cline{13-13} \cline{14-14} \cline{15-15} \cline{16-16} \cline{17-17}
Naive & 1.25 & 1.25 & 3.09 & 3.15 & 2.2 & 2.18 & 5.52 & 5.51 & 1.25 & 1.23 & 3.11 & 3.15 & 2.19 & 2.21 & 5.46 & 5.4\\
Regression & 0.41 & 0.33 & 0.41 & 0.31 & 1.01 & 0.69 & 0.96 & 0.66 & 1.23 & 0.47 & 3.02 & 0.52 & 2.18 & 1.15 & 5.39 & 1.18\\
IPW & 0.66 & 0.68 & 1.67 & 1.62 & 1.08 & 1.03 & 2.63 & 2.6 & 0.99 & 0.96 & 2.4 & 2.46 & 1.46 & 1.48 & 3.61 & 3.56\\
AIPW & 0.28 & 0.23 & 0.3 & 0.21 & 0.65 & 0.4 & 0.6 & 0.4 & 0.99 & 0.4 & 2.42 & 0.45 & 1.48 & 0.85 & 3.74 & 0.9\\
\hline
BalanceHD (weights only) & 0.35 & 0.37 & 0.84 & 0.86 & 1.03 & 0.97 & 2.46 & 2.44 & \textbf{0.79} & 0.76 & \textbf{1.89} & 1.95 & 1.27 & 1.28 & 3.09 & 3.09\\
BalanceHD & \textbf{0.19} & \textbf{0.18} & \textbf{0.22} & \textbf{0.15} & 0.63 & 0.37 & \textbf{0.59} & 0.38 & 0.81 & \textbf{0.33} & 1.98 & \textbf{0.38} & 1.3 & 0.76 & 3.26 & 0.83\\
\hline
IPW-d(0) & 0.61 & 0.62 & 1.55 & 1.49 & 0.74 & 0.68 & 1.75 & 1.68 & 0.96 & 0.9 & 2.29 & 2.34 & 1.36 & 1.02 & 3.29 & 2.53\\
AIPW-d(0) & 0.28 & 0.21 & 0.3 & 0.2 & 0.64 & 0.33 & 0.64 & 0.33 & 0.97 & 0.37 & 2.34 & 0.43 & 1.4 & 0.67 & 3.5 & 0.77\\
IPW-d(-1) & 0.59 & 0.59 & 1.48 & 1.41 & \textbf{0.62} & 0.54 & 1.35 & 1.31 & 0.84 & 0.86 & 1.98 & 2.26 & \textbf{1.09} & 0.85 & \textbf{2.6} & 2.12\\
AIPW-d(-1) & 0.28 & 0.21 & 0.3 & 0.19 & 0.64 & \textbf{0.32} & 0.65 & \textbf{0.31} & 0.86 & 0.35 & 2.08 & 0.42 & 1.14 & \textbf{0.61} & 2.88 & \textbf{0.72}\\
\hline
\end{tabular}

\caption{\textbf{Root Mean Squared Error (RMSE) for different data generating models and estimation methods.}  We regularize outcome and propensity score models using either ridge penalty (R) or lasso penalty (L), consider higer signal-to-noise (SNR) regimes, $S_Y = 5$, and lower signal-to-noise regimes, $S_Y = 2$.  We also consider high overlap setting, $S_T = 2$ and low overlap settings $S_T = 4$.  Finally, we consider lower dimensional problems ($p=100$) and higher dimensional problems ($p=1000$).  In all simulations $n=500$. Deconfounding outperforms BalanceHD in low overlap situations. Results correspond to the average of 100 runs. \label{tab:rmse}}
\end{table*}

\subsection{Results}

\paragraph{Overall RMSE} In Table \ref{tab:rmse} we report the root mean squared error of the ATT, for ten different estimators under several different data generating and inferential mechanisms.  We focus in particular on two special cases of the deconfounding score: IPW-(0)/AIPW-d(0), for which $e(X)$ is replaced with $e_d(X)$ for a $d(X)$ that is equally correlated with the prognostic and propensity scores, and IPW-d(-1)/AIPW-d(-1) for which we replace $e(X)$ with $E[e(X)\mid m_0(X)]$.

In all simulation settings in Table \ref{tab:rmse}, methods that use alternative weighting schemes have the lowest RMSE (either balanceHD or our deconfounding score methods).
In high overlap regimes, balanceHD dominates; this was expected becuase the method was optimized for this setting.
However, in low overlap cases our deconfounding score methods tend to have the lowest error, especially in the higher dimensional regime.

Among the deconfounding score estimators, the estimators that set $w = -1$ appear to dominate, with AIPW-d(-1) winning when the appropriate lasso regularization is used, and IPW-d(-1) winning when the mismatched ridge regularization is used.
Recall that these correspond to estimators that use the prognostic score both in the regression estimator and to construct weights.
It is somewhat surprising that these estimators outperform the regression estimator, because they all appear to be making use of the same information.
Although the $w = 0$ setting appears less effective in this set of experiments, we note that it almost always represents an improvement on the standard IPW and AIPW estimators.
In addition, we suspect that there are cases where the propensity score can be estimated with far greater certainty than the outcome model (e.g., in field experimentation contexts where the propensities are known) where the $w=0$ case could prove favorable.

\begin{figure}[h!]
  \begin{center}
    \includegraphics[width=0.48\textwidth]{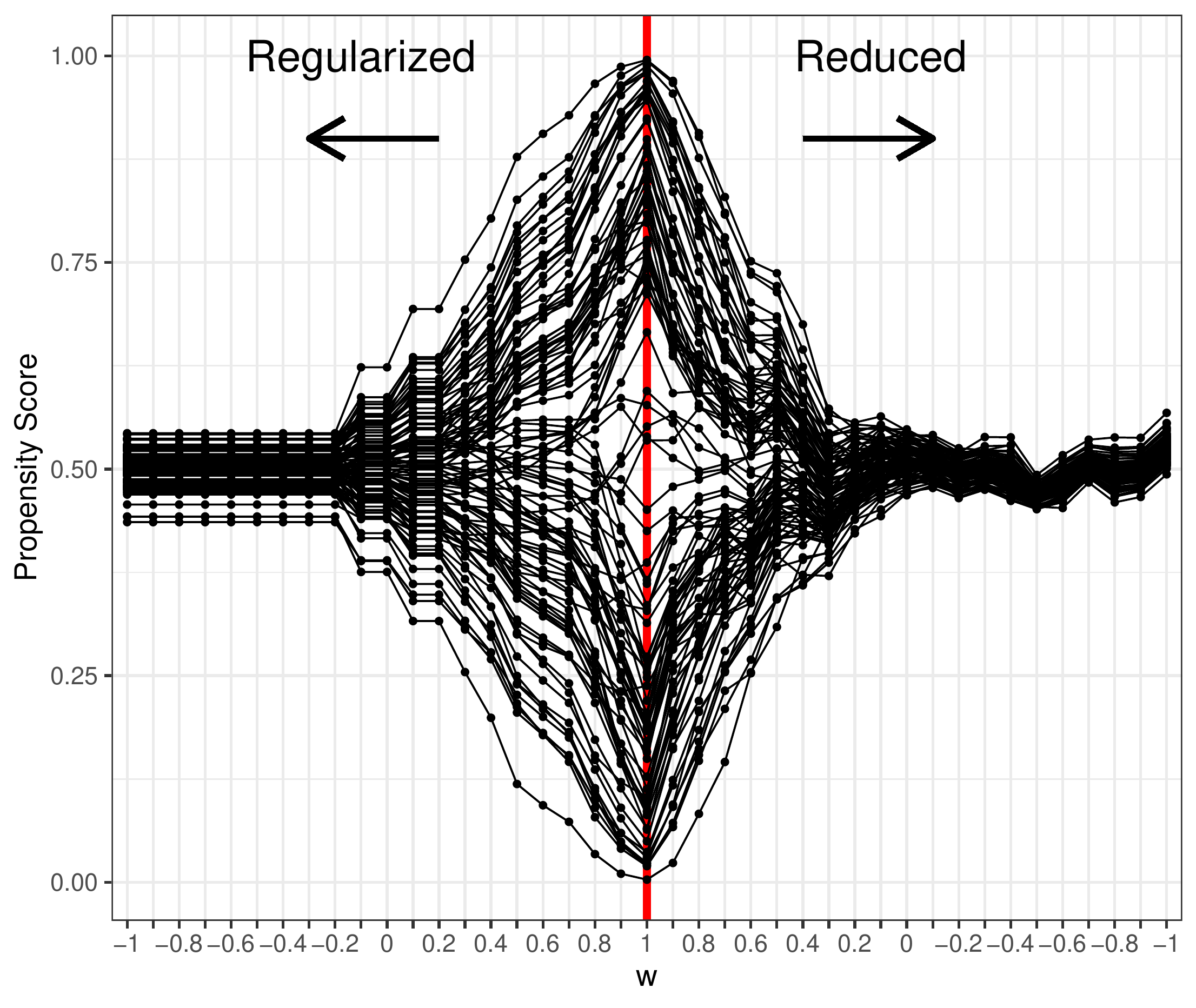}
  \end{center}
\caption{\textbf{Shrinkage plot demonstrating 
    of the per-unit propensity scores as a a function of $w$ for both classical regularized propensity scores and our deconfounding scores.}  The ridge penalty for the regularized estimates are chosen to match the variance of the propensity score from the deconfounding scores, with the variance decreasing as  $w_2$ decreases to -1. However, as shown in Figure \ref{fig:bias_var}, classical shrinkage incurs bias, whereas the deconfounding score typically does not.}
    \label{fig:shrinkage_plot}
\end{figure}

\paragraph{Comparison to Regularization}
To understand how deconfounding scores can improve RMSE relative without inducing bias, we compare the effect of lowering the parameter $w$ (that is, moving the deconfounding score away from the propensity score) to directly regularizing or propensity score estimates.  
We depict a bias variance decomposition of the MSE for each of these strategies in Figure~\ref{fig:bias_var}.
For each value of $w$ on the horizontal axis, we choose the propensity score regularization or clipping parameter so that the variance of the propensity scores used in AIPW and IPW matches the variance of $e_d(X)$ at each value of $w_2$ (See Figure \ref{fig:shrinkage_plot}).
In Figure \ref{fig:bias_var}, it is evident that the RMSE of the classical IPW and AIPW estimators increases with additional regularization
.
Here, although additional regularization of the propensity score has moderate effects on variance, the dominating factor is bias, which increases with additional regularization.
In contrast the deconfounding scores strongly reduce bias as a function of $w$.
As a benchmark, we also include an ``oracle'' IPW deconfounding score in which $e_d(X)$ is computed using the true prognostic score $m_0(X)$ and propensity score $e(X)$.

\section{Discussion}

In this work, we introduced the notion of general deconfounding scores, showed how they can be computed from data, and demonstrated as a proof-of-concept that they can be used to construct useful weighting estimators in a simple setting.

A number of directions remain to flesh out this work.
First, there remain major gaps in the theory about how employing deconfounding scores affects the asymptotic and finite-sample distributions of estimators.
It would be particularly interesting to understand how deconfounding score estimation might affect asymptotic semiparametric variance calculations.
Secondly, we have only worked out how to compute deconfounding scores in a highly simplified setting.
Some of our results may be directly extensible via latent Gaussian models, especially with VAE-like representations.
It could also be interesting to turn the bias formula we presented in \eqref{eq:empirical bias} into a penalty for representation learning.
To this end, it is worth noting that the bias could also be estimated using the same efficient semiparametric estimation approaches that are applied to the ATE.

Finally, we highlight the surprising performance of using the prognostic score twice---once for regression, and once for weighting.
In our simulations, this turned out to be a highly effective approach, even when the outcome model was strongly biased by regularization.
It is unclear whether this insight generalizes beyond the simple setting that we tested, but this question could be pursued independently of the (much more convoluted) topic of deconfounding scores.

\bibliographystyle{icml2020}
\bibliography{reducer}
\vfill

\input{appendix}

\end{document}

%% file: appendix.tex
\newpage
\appendix
\section{ATT Definitions}
Here we present the analogs of the IPW (3) and AIPW (4) estimators of the ATE for the average treatment effect among the treated (ATT):
\begin{align}
\label{eq:IPW_ATT}
\hat \tau_{IPW}^{ATT} := \hat E\left[\left(T - (1-T)\frac{\hat e(X)}{(1-\hat e(X))}\right)Y\right],
\end{align}
and 
\begin{align}
\label{eq:AIPW_ATT}
\hat \tau_{AIPW}^{ATT} :=  %
\hat E[Y \mid T = 1] - \hat E[\hat m_0(X) \mid T = 1] +
\hat E\left[-(1-T)\left(\frac{\hat e(X)}{(1-\hat e(X))}\right)(Y - \hat m_0(X))\right].
\end{align}

\section{Proof of Proposition 1}
 Note that $\tau = \mu^{(1)} - \mu^{(0)} = E[E[Y(1)|d(X)]] -  E[E[Y(0)|d(X)]]$

 \begin{align}
 E\left[\frac{TY(1)}{e_d(X)}\right] &= E\left[\frac{E[TY(1)|e_d(X)]}{e_d(X)}\right]\\
 &= E\left[\frac{\Cov(T, Y(1) | d(X))}{e_d(X)}+\frac{E[T|d(X)]E[Y|d(X)]}{e_d(X)}\right]\\
 &= E_P\left[\frac{\Cov(T, Y(1) | d(X))}{e_d(X)}\right] +E\left[E[Y(1)|d(X)]\right]\\
 & =  E_P\left[\frac{\Cov(T, Y(1) | d(X))}{e_d(X)}\right] + \mu^{(1)}
 \end{align}
 Similarly,
 \begin{align}
 E\left[\frac{(1-T)Y(0)}{1-e_d(X)}\right] &= E_P\left[\frac{\Cov((1-T), Y(0) | d(X))}{(1-e_d(X)))}\right] + \mu^{(0)}
 &=-E_P\left[\frac{\Cov(T, Y(0) | d(X))}{(1-e_d(X)))}\right] + \mu^{(0)}.
 \end{align}

 And thus,
 $\tau_d^{ATE} = \mu^{(1)} - \mu^{(0)} + E_P\left[\frac{\Cov(Y(1), T \mid d(X))}{e_d(X)} + \frac{\Cov(Y(0), T \mid d(X))}{1-e_d(X)} \right]$.

 \section{Proof of Theorem 1: Hyperbola Derivation}

The constraint in (11) is equivalent to

 $$\gamma^T \left(\frac{\alpha \beta' + \beta \alpha'}{2}\right)\gamma = \alpha'\beta$$

 If $\alpha$ and $\beta$ are not colinear then $\left(\frac{\alpha \beta' + \beta \alpha'}{2}\right)$ is rank 2 with exactly one positive eigenvalue and one negative eigenvalue.  
 Specifically, the eigenvectors $[u_1, u_2]$ and eigenvalues $\lambda_1$ and $\lambda_2$ are 
    
 \begin{align}
 u_1 &= \frac{(\alpha + \beta)}{\sqrt{2 + 2\alpha'\beta}}, u_2 = \frac{(\alpha - \beta)}{\sqrt{2 - 2\alpha'\beta}}\\
 \lambda_1 &= \frac{(\alpha'\beta+1)}{2}, \lambda_2 = \frac{(\alpha'\beta-1)}{2}
 \end{align}

 Then we have a solution characterized by the hyperbola
 \begin{equation}
 \label{eq:hyperbola constraint}
 \frac{\alpha'\beta + 1}{2\alpha'\beta}\left(\frac{(\alpha + \beta)'\gamma}{\sqrt{2 + 2\alpha'\beta}}\right)^2 + \frac{\alpha'\beta - 1}{2\alpha'\beta}\left(\frac{(\alpha - \beta)'\gamma}{\sqrt{2 - 2\alpha'\beta}}\right)^2 = 1
 \end{equation}
 Simplifying notation by collapsing scalars into $w_1$ and $w_2$ yields the result.

\section{Computing $e_d(X)$ in the Gaussian Case}
To derive $e_d(X)$ we note that $e_d(X) = E[T \mid d(X)] = E[e(X) \mid d(X)]$.  We then write $\beta$ as a linear combination of $\gamma$ and an orthogonal component, i.e $\beta = m \gamma + \sqrt{1-m^2}z$ where $\gamma'z = 0$ and $m \in [0, 1]$, so that (9) can be written as $e(X) = \text{logit}^{-1}((\beta_0 + s_Tmd(X) + s_T\sqrt{1-m^2}X'z))$ Consequently, 
\begin{align}
\label{eqn:t_given_d}
E[T\mid d(X)] = 
&E[\text{logit}^{-1}(\beta_0 + s_Tmd(X) + s_T\sqrt{1-m^2}X'z) \mid d(X)].
\end{align} 
Since $X\sim N(0,I)$ and $z$ is a random unit vector, $X'z \sim N(0, 1)$ .  We can easily construct an accurate Monte Carlo estimator $E[T\mid d(X)]$ by averaging \eqref{eqn:t_given_d} over several standard normal draws of $X'z$.